%% file: RJwrapper.tex
\documentclass[a4paper]{report}
\usepackage[utf8]{inputenc}
\usepackage[T1]{fontenc}
\usepackage{RJournal}
\usepackage{amsmath,amssymb,array}
\usepackage{booktabs}

\usepackage{longtable}

\makeatletter
 \let\@cite@ofmt\@firstofone
 \def\@biblabel#1{}
 \def\@cite#1#2{{#1\if@tempswa , #2\fi}}
\makeatother
\newlength{\cslhangindent}
\newlength{\csllabelwidth}
 {\begin{list}{}{%
  \setlength{\itemindent}{0pt}
  \setlength{\leftmargin}{0pt}
  \setlength{\parsep}{0pt}
  \ifodd #1
   \setlength{\leftmargin}{\cslhangindent}
   \setlength{\itemindent}{-1\cslhangindent}
  \fi
  \setlength{\itemsep}{#2\baselineskip}}}
 {\end{list}}
\usepackage{calc}

\begin{document}

\sectionhead{Contributed research article}
\volume{XX}
\volnumber{YY}
\year{20ZZ}
\month{AAAA}

\begin{article}
  \input{pandemonium}
\end{article}

\end{document}

%% file: pandemonium.tex
\title{\texttt{pandemonium}: High Dimensional Analysis in Linked Spaces}

\author{by Gabriel McCoy, German Valencia, and Ursula Laa}

\maketitle

\abstract{%
A common challenge in data analysis is uncovering relationships between predictors and responses in problems involving large numbers of both. When the number of predictors and responses is limited, visual approaches are particularly effective. We present an R package, pandemonium, designed to explore such problems by combining cluster analysis with linked visualisations. Clustering is performed in one set of variables to identify regions with similar patterns in that space. The resulting clusters are simultaneously visualised in both spaces using linked views based on non-linear dimension reduction and animated tours. We introduce the package through two examples that illustrate different types of linked spaces. In the first example, we consider how a set of input variables is mapped to latent activations in a neural network regression model, to identify input combinations that result in similar activation patterns. In the second example, we analyse a complex multivariable mathematical model arising in physics to investigate how structure in the predictor space relates to the responses.
}

\section{Introduction}\label{introduction}

Many real-world phenomena are characterised by high levels of complexity and are naturally described using multiple predictors and multiple responses. Such problems give rise to two high-dimensional spaces -- often referred to as input and output spaces -- that are intrinsically linked. The nature of this linkage can vary considerably. In some cases, it is explicitly defined, for instance, through a system of mathematical equations or a computational model that maps predictors to responses. These can be considered input/output models and an overview of different approaches in \emph{visual parameter space exploration} in the visual analytics literature is given in \citet{vpse}. In other situations, the relationship between the two spaces is not known a priori, as in the analysis of multivariate observational data, where the goal is to uncover previously unknown dependencies, patterns, or structures linking the variables. One example is comparing ground-based measurements with remote sensing data in ecological studies, where we might aim to predict costly measurements from remote sensing data.

Regardless of how the connection between these spaces is defined, gaining insight into their structure and interdependence is challenging due to their dimensionality and complexity. Visual inspection enables analysts to formulate intuitive hypotheses, identify patterns, detect anomalies, and guide subsequent quantitative analyses. However, traditional visualisation techniques often focus on a single space at a time, making it difficult to reason about relationships that span across both domains.
The package presented in this paper addresses this limitation by providing an interactive visual framework for the joint exploration of two linked high-dimensional spaces. The interface allows users to import data and flexibly assign variables to each space, accommodating a wide range of problem formulations. Both spaces are visualised simultaneously and are dynamically linked, enabling changes or selections in one space to be immediately reflected in the other.

To support targeted exploration, the tool uses clustering within one of the spaces, allowing users to select meaningful subsets of points or subspaces based on similarity in that space. These selections serve as a lens through which the corresponding structure in the second space can be examined.
The work builds on a prototype that was designed for the exploration of connections between parameters of models in theoretical particle physics, and the model predictions for a range of experimentally observable quantities \citep{Laa:2021dlg}. We use \CRANpkg{shiny} \citep{shiny} to enable interactive selection of cluster settings by the user, and to provide linked interactive graphics to study the connected spaces.

Related work in R generally focuses on the interactive exploration of clustering results, for example, through interactive dendrograms \citep{JSSv076i10}, combining clustering with dimension reduction \citep{drtool} or the use of interactive graphics for the refinement of clustering solutions in a spin-and-brush approach \citep{lionfish}.
Our contribution is to define a generic framework that allows the user to explore the two connected spaces while interactively changing settings for the hierarchical clustering used to group the observations. This enables an in-depth investigation of connections between the spaces, and is broadly applicable beyond the original use-case in particle physics.

\section{Notation}\label{notation}

To define the interface generically, we introduce a standard notation that will be used to explain the tool, as well as in the different applications below.

We consider a set of \(n\) observations for which we have information in two separate spaces. Each space is defined as a subset of variables in the full data where a natural separation typically occurs. For example, the original inputs and the latent space of a statistical model would constitute two such spaces. Here, the naming of these spaces will be based on the usage within the analysis framework: a \emph{clustering space} (input for the cluster analysis) and a \emph{linked space} in which the cluster results are also explored.
We denote variables in the clustering space \(Y_i, i=1,...,p_c\), where \(p_c\) is the number of variables assigned to the clustering space. \(Y_{ai}\), \(a=1,...,n\) then corresponds to the value of the clustering variable \(i\) for the observation \(a\). Similarly, observed values for variables in the linked space are denoted by \(X_{bj}, b=1,...n, j=1,...,p_l\), with \(p_l\) the number of variables assigned to the linked space. For each space, a variance-covariance matrix \(\Sigma_Y\) or \(\Sigma_X\) may also be provided, and this information can be used for the definition of coordinates used in the calculations. Equivalently, one of these two matrices is used to define the metric in the clustering space. Finally, for each observation \(c\) we might have \(p_k\) variables with additional information, \(Z_{ck}, k=1,...,p_k\), which are not assigned to either space; instead, they might be used to define groups external to the app, additional coordinates, or a score for each observation. Within \CRANpkg{pandemonium}, we can perform comparisons between clustering results and the additional information.

For each of the two spaces, the user can select a coordinate transformation that converts the raw input variables into a coordinate representation, denoted \(\widetilde Y\) and \(\widetilde X\). This is particularly important in the clustering space, where distances between observations will depend on the coordinate transformation selected. Moreover, distances will be relevant in both spaces for visualisations using dimension reduction. We denote by \(d_Y(a,a')\) and \(d_X(a,a')\) the distance between the two observations \(a, a'\) in the corresponding spaces, where the coordinate space representation is used in the distance calculation.

A simple choice for the coordinate transformation would be to standardise the variables to have a mean value of zero and a standard deviation of one. Coordinate functions can also take into account information in the relevant variance-covariance matrix and use values stored in the \(Z_k\), an example of which is given for the second case study below.

Finally, additional information can be added in the form of \emph{scores}, which can be displayed in the app. The scores might be computed based on information in the \(Y_i\), \(X_j\) and \(Z_k\), where \(Z_k\) may be used to directly input external information, for example, model residuals computed based on predicted and observed values (see example in the first case study).

\section{App design}\label{app-design}

\begin{figure}

{\centering \includegraphics[width=1\linewidth]{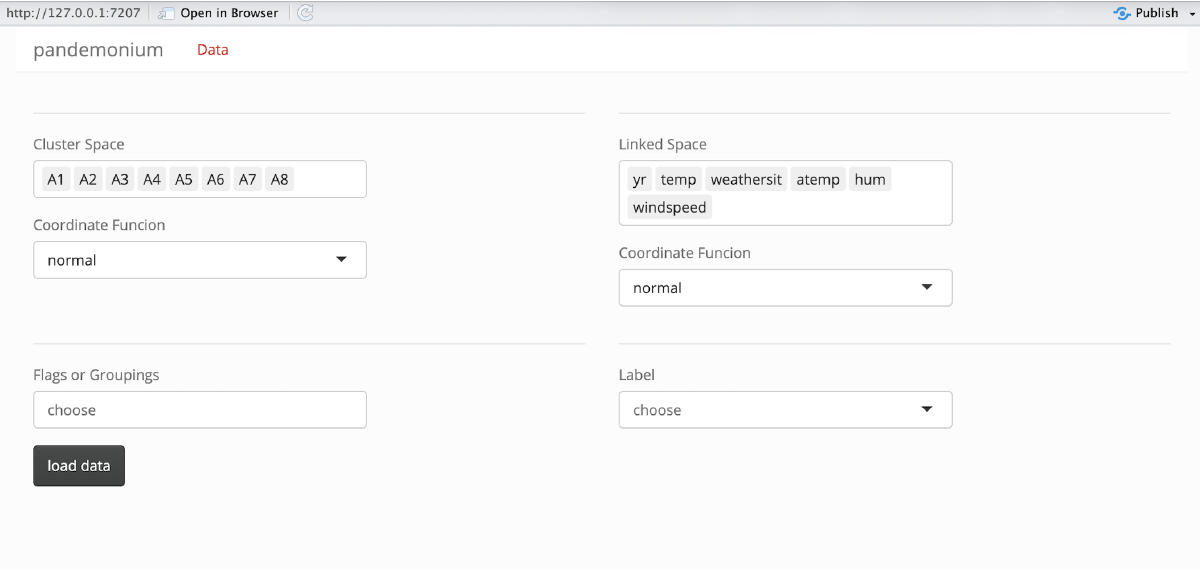} 

}

\caption{The Data page shown here is used to input the data provided to pandemonium into each field.}\label{fig:data-page-image}
\end{figure}

The \CRANpkg{pandemonium} app is structured into two main sections. The first section, seen in Fig. \ref{fig:data-page-image}, is a data input page that allows the user to set up the analysis starting from raw data inputs, while the analysis is performed in the second section shown in Fig. \ref{fig:input-tab-image}.

\subsection{Data input page}\label{data-input-page}

All numeric variables can be placed into either the clustering space or the linked space. One variable can be selected as a data entry ID or label and all variables can be used to define groupings within the data. If multiple variables are selected in the grouping, they will be combined into all combinations occurring in the data. Due to limitations in colour palettes, a maximum of 13 groups can be distinguished in the displays.
Grouping is particularly helpful for exploring categorical variables and how they cluster in the two spaces. A comparison of external grouping with the clustering result may also be of interest.

A first processing step will translate the raw data into a coordinate space, where a transformation is defined via coordinate functions that can be selected separately for each space. Coordinate functions can be defined by the user and passed to \CRANpkg{pandemonium} when launching the app, or the user can select from a set of predefined functions. A simple example would be to centre and scale each variable, while the physics use-case described below uses the variance-covariance matrix for the definition of coordinates.

Additional pre-processing is available for the information in \(Z\), where users can define a score function that defines how to break the input into groups, and there is limited support for handling missing values, which can be imputed using \CRANpkg{VIM} \citep{VIM}. Examples below use quantiles of model residuals or \(\chi^2\) values from a fitted model as score functions.

After finishing the pre-processing selection, the user can launch the corresponding analysis page by clicking the \emph{load data} button.

\subsection{Analysis page}\label{analysis-page}

\begin{figure}

{\centering \includegraphics[width=1\linewidth]{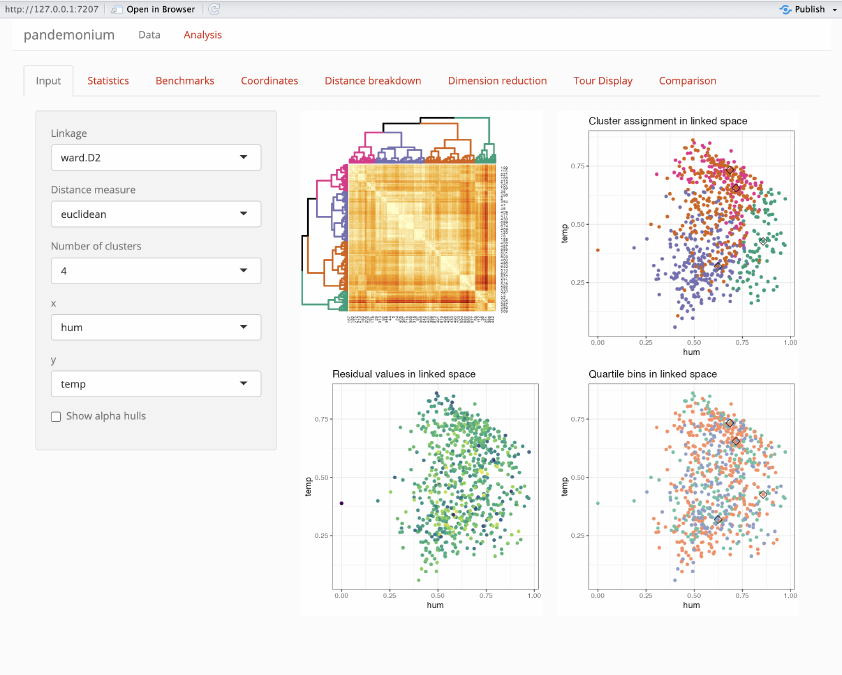} 

}

\caption{The analysis page contains eight tabs, shown here is the input tab where options for the hierarchical clustering can be selected interactively, and a first overview of the results is shown.}\label{fig:input-tab-image}
\end{figure}

The data prepared in the data input page provides the basis for the analysis, combining clustering with high-dimensional visualisation techniques. To provide the required flexibility in a clear structure, the analysis page is broken up into eight tabs, as shown in Fig. \ref{fig:input-tab-image}.

\subsubsection{Clustering setup and parameter selection}\label{clustering-setup-and-parameter-selection}

The starting point is to cluster the observations using hierarchical clustering on the coordinate presentation of the clustering space, \({\widetilde {Y}}\). Clustering options can be selected in the initial input tab, which also shows a first set of visualisations to guide the parameter selection.
\CRANpkg{pandemonium} uses hierarchical clustering for a repeatable clustering result, which allows for increasing or decreasing the number of clusters in a nested fashion. The user can select the distance metric, cluster linkage and number of clusters used from the sidebar. With each new selection, the clustering results and the displays are updated. While supplying a new distance metric function is currently not possible, the user can provide explicit distance values that can be used in \CRANpkg{pandemonium} when launching the app.

To give an overview of the results, the input tab shows a heatmap with the clustering dendrogram as obtained from \CRANpkg{dendextend} \citep{dendextend}, as well as projected views of the linked space shown with two user-selected variables. These plots will show the scores and bins, if defined, as well as the clusters. Optionally, convex hulls can be shown with the cluster results; this is done using the \CRANpkg{alphahull} \citep{alphahull} package, with an alpha value selected via the interface.

Additional guidance for selecting the preferred number of clusters is available in the \emph{Statistics} tab, which shows a set of cluster statistics computed for the current selection, for up to eight clusters.
The set of available clustering statistics uses some of those implemented in the \CRANpkg{fpc} \citep{fpc} package, together with a characterisation in terms of cluster size and distances between cluster benchmarks. Here a benchmark point for each cluster \(C\) is defined as the observation \(c\) that minimizes
\begin{equation}
f(c, C) = \sum_{a \in C} d_Y(c, a)^2.
\label{eq:bp}
\end{equation}
Based on the benchmark observations, we define the corresponding cluster radius as the maximum distance between any point in the cluster and its benchmark:
\begin{equation}
r_C = \max_{a \in C} d_Y(c, a).
\label{eq:rc}
\end{equation}
Finally, the cluster diameter is defined as the largest distance between any two observations that are assigned to the same cluster.

\subsubsection{Summaries of the clustering results}\label{summaries-of-the-clustering-results}

The next few tabs should give an overview of the clustering results, primarily focusing on exploring the clustering in the linked space.

First, the benchmark tab gives the coordinate values in the linked space for the cluster benchmark points (as defined from Eq. \eqref{eq:bp}). Additionally, the cluster radius and diameter are also included, as well as the score when a score function is defined. If other groupings are provided in the input page, a second section shows the corresponding information for the group benchmarks.

A more detailed view of the clustering in the linked space is provided in the coordinate tab. At the top of the page, a set of scatterplots in the two variables selected for display (here, the selection from the input tab is used) shows the coordinate values in the clustering space as a colour gradient. This enables the detection of general patterns between the two spaces. Below, an interactive parallel coordinate plot shows the individual observations in the clustering space, with benchmark observations highlighted. The user can select if coordinates should be centred and/or scaled for the display, and individual clusters can be removed from the display for a more focused analysis of selected groups. By combining information in the two displays, we can get an idea of which variables in the clustering space are most important for separating the clusters, and how these translate to patterns in the linked space.

The distance breakdown tab shows a histogram of within and between cluster distances for each cluster, to compare cluster separation. Similar to silhouette widths \citep{ROUSSEEUW198753}, this can be used to confirm the clustering of the data. For reference, the overall distance histogram is always drawn in the background.

\subsubsection{Linked high-dimensional data visualisation}\label{linked-high-dimensional-data-visualisation}

With the clustering determined, these results are used to explore the spaces using linked high-dimensional visualisations. Since they provide complementary insights \citep{Lee_Laa_Cook_2022}, we use both non-linear dimension reduction and tour methods \citep{tourreview}, i.e.~animated linear projections, in linked displays using \CRANpkg{crosstalk} \citep{crosstalk}.
These visualisations allow both spaces to be shown side by side or each space to be shown with different options. Points can be coloured by clustering, grouping, scores or bins. Brushing results in highlighting of the selected points across all displays. Note that the display has to be initialised first for the linked brushing to take effect.

\begin{figure}
\includegraphics[width=1\linewidth]{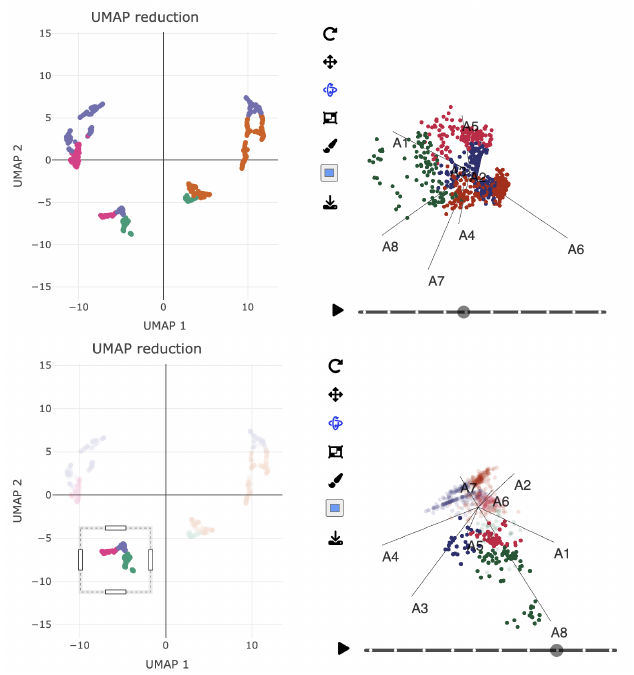} \caption{Linking visualisation of the clustering space with UMAP (left) and tour (right). The colours show the result from hierarchical clustering, and we can use brushing to understand where differences in this grouping come from that are not found with UMAP. The bottom panel shows one group identified in the UMAP view, brushed and highlighted in the tour view.}\label{fig:crosstalk-ex-pdf}
\end{figure}

The complementarity of clustering, non-linear dimension reduction and tour visualisations is illustrated in Fig. \ref{fig:crosstalk-ex-pdf}: groups identified through hierarchical clustering are shown in colour and do not match the grouping found in the UMAP visualisation. By exploring the UMAP clusters through linked brushing with the tour visualisation, we can gain some understanding of where the differences come from.

For example, we brush the UMAP cluster containing three different colours, and look at this group of observations in the tour display. This group is scattered widely across the space, and there are no gaps between the clusters. If instead we focus on the UMAP cluster containing only pink and purple observations, we see that these extend along different linear directions in the full space, but are overlapping near the centre.

The dimension reduction tab shows two side-by-side dimension reduction plots. Options are provided for choosing which space is shown in each plot, what grouping each plot uses and the dimension reduction technique used to create the plot. Default options are t-SNE \citep{Rtsne2} and UMAP \citep{2018arXiv180203426M}, additional methods can be provided by the user when launching the app. Note that the default options use default values for all tuning parameters; to specify different values, requires providing the corresponding function with the desired values for the parameters.

The tour tab also shows a side-by-side display of two tour views, where again both the space shown and the grouping used for colouring can be selected for each display separately. In addition, the user can select between different tour display options, including a slice tour display \citep{tourr-slice}, and different options for the tour path generation. Of particular interest is a guided tour that optimises for separation between the groups in the final linear projection.

The linking between the displays spans across the two tabs, such that the user can investigate groups of points identified in any one of the displays across all other non-linear dimension reduction and tour views.

\subsubsection{Making comparisons}\label{making-comparisons}

Often, we also wish to compare clustering results using different settings. This is possible in the comparison tab, where two different clustering settings can be selected. The overview visualisations provided in the input tab are shown for both results, and an additional heatmap display shows the overlap between clusters in the two solutions.

\section{Implementation}\label{implementation}

\subsection{Overall structure}\label{overall-structure}

We use \CRANpkg{shiny} \citep{shiny} to implement the \CRANpkg{pandemonium} GUI. The main function \texttt{pandemonium()} is used to launch the app, and its arguments can be used for the customisation of GUI options, including modular inputs for some of the functions used within the app. All functions that compute results specific to the app, as well as those used to generate the different graphs, are structured into a set of scripts to keep the \texttt{server()} definition as simple as possible. All interactive graphs are linked via \CRANpkg{crosstalk} \citep{crosstalk}, to explore subsets with linked brushing across the various tabs.

The \CRANpkg{pandemonium} package also provides the functions \texttt{makePlots()} and \texttt{writeResults()} to allow the user to reproduce results and graphics identified in the GUI outside of the interactive interface. The underlying functions used to generate the results are the same as those used in the app whenever possible, to ensure consistent results. An exception is the generation of tour outputs, where a different output format is desirable for exported results. To reproduce the results, the user needs to provide the corresponding settings selected in the app in list format in the \texttt{settings} argument. The corresponding item names and allowed values are documented in a vignette in the package.

\subsection{Implementation of the tour displays}\label{implementation-of-the-tour-displays}

Tour views are made using \CRANpkg{detourr} \citep{detourr, RJ-2023-052} and allow for many different types of tours to be used through \CRANpkg{tourr} \citep{tourr}. After selecting all settings, the user needs to press \emph{build tour} to launch the computation. This is done to save the tour from recomputing when any update is made in the reactive chain, since this can take a long time particularly for guided tours \citep{CBCH94}. By saving the entire tour path, we can also easily access information from the tour like the final projection of a guided tour. Such a selected projection can be used as the starting projection for a radial tour \citep{2022arXiv221005228L}, or we can use the full path for showing the same tour side by side with different groupings mapped to colour, for a detailed comparison of the grouping patterns across the space.

A major advantage of using \CRANpkg{detourr} is the generation of interactive 2D and 3D tour visualisations, in particular also allowing for linked brushing with other displays. From the precomputed tour path, the final frame can be accessed using the \emph{hold frame} button in the GUI. Using the same tour path with different colouring is done by using the \emph{copy next} tour button in the GUI.

Within the GUI, the tour display can be changed between a regular scatterplot display and a slice tour display \citep{tourr-slice}. This is achieved by defining a function for the display as a reactive variable, which switches between \texttt{detourr::show\_slice()} and \texttt{detourr::show\_scatter()} based on the user's selection. The user can also select the dimension of the tour, some guided tour index functions are not defined for 3D projections, therefore the server updates the UI to remove these options using \texttt{shiny::updateSelectInput()}. Other options that are only required if other settings are selected are hidden using \texttt{shiny::conditionalPanel()} within the UI.

Within the \texttt{makePlots} function, the user instead provides all options in the function call. In this case, \CRANpkg{detourr} is used to both calculate the tour path and create the output. An additional option in the \texttt{makePlots} function is to produce a static plot. This takes the final frame of the tour and produces a \CRANpkg{ggplot2} output instead of a \CRANpkg{detourr} output. This is particularly useful to show the optimal projection according to a projection pursuit index used with a guided tour.

\subsection{Linking with crosstalk}\label{linking-with-crosstalk}

A key feature of the app is the ability for linked brushing between tour and dimension reduction displays, to further explore the clustering and linked spaces. To achieve this, the tour displays are created using \CRANpkg{detourr} \citep{detourr} and the dimension reduction plots generated as static interactive visualisations using \CRANpkg{plotly} \citep{plotly}. Both of these are built using \CRANpkg{htmlwidgets} \citep{htmlwidgets}, which allows for linked brushing between plots using \CRANpkg{crosstalk} \citep{crosstalk}.

To set up linked brushing, the data for both the tour and the dimension reduction plot is combined into a single data frame. This data frame is passed to a \CRANpkg{crosstalk} object used in all displays. This maintains the reactive chain in the \CRANpkg{shiny} \citep{shiny} app but results in all of the plots needing to be updated when any of the data is changed. The setup is sketched in the following code:

\begin{verbatim}
# setting up the shared data frame in a reactive value
coordinates_data <- shiny::reactive({
  # coordinates of selected space, dimension reduction output
  cbind(rv$mat,rv$dimred$Y)
})

# enabling the sharing with crosstalk
shared_data <- crosstalk::SharedData$new(coordinates_data)

# implementation of the linking with detourr
output$detour <- detourr::shinyRenderDetour({
  # input data is the shared crosstalk object
  # projection is entire space selected
  detourr::detour(shared_data, 
                  detourr::tour_aes(projection = rv$matnames, 
                                    colour = rv$colour)) %>%
    detourr::tour_path(tourr::grand_tour(), fps = 60, max_bases=20) %>%
    detourr::show_scatter(alpha = 0.7, axes = TRUE, palette = rv$pal)
  })

# implementation of linking with plotly
output$dim_red <- plotly::renderPlotly({
  # input data is the shared crosstalk object
  plotly::plot_ly(shared_data, x= ~dim1, y= ~dim2, 
                    colour = rv$colour, colours = rv$pal) %>%
    plotly::add_trace(type = "scatter", mode = "markers") %>%
    plotly::layout(xaxis = list(scaleanchor = y), showlegend = FALSE) %>%
    # for linked brushing we want to highlight selected points
    plotly::highlight(on = "plotly_slected", off = "plotly_deselect")
})
\end{verbatim}

\subsection{Modular inputs}\label{modular-inputs}

To maintain some flexibility within the GUI \CRANpkg{pandemonium} was implemented in a modular fashion, such that a range of computations can be replaced by user-defined functions. This includes allowing alternative methods for dimension reduction, as well as user definitions for coordinate and score functions. These functions can be provided as a named list when calling \texttt{pandemonium()} to launch the app. The name will be added to the selection menu in the app, which will then call the corresponding function for the computation.

By default, \CRANpkg{pandemonium} provides two options for dimension reduction, tSNE via the \CRANpkg{Rtsne} \citep{Rtsne1, Rtsne2} package and UMAP via the \CRANpkg{uwot} \citep{uwot} package. To define an alternative, the provided function should be written as illustrated below.

\begin{verbatim}
lleFun <- function(mat, ...) {
  ret <- list()
  ret$Y <- Rdimtools::do.lle(mat)$Y
  ret
}

pandemonium(df, dim_reduction = list(lle = lleFun))
\end{verbatim}

Note that internally, when calling the function, two inputs are provided: \texttt{mat}, the coordinate representation of all observations, and \texttt{dist}, the pairwise distance matrix. These should be expected inputs of the provided function, though \texttt{...} may be used for the input not needed in the computation. The function return value should provide the two-dimensional NLDR representation in the \texttt{\$Y} component, either in matrix or data frame format.

In the current version, \CRANpkg{pandemonium} does not allow for interactive selection of hyperparameters for the dimension reduction. The available methods use default parameters, while for user input the hyperparameters should be fixed in the definition.

Similarly, functions can be defined for score and coordinate definition; details are given in the package vignettes. When calling the app, the user-defined functions are passed in as:

\begin{verbatim}
pandemonium(df, getScore = chi2Score, 
              dim_reduction = list(tSNE = tSNE, umap = umap), 
              getCoords = list(normal = normCoords))
\end{verbatim}

\section{Usage}\label{usage}

The simplest way to launch the app is with all the clustering space, linked space and additional data in a single data frame passed in the \texttt{df} parameter as \texttt{pandemonium(df)}. With this, the numeric variables can be selected into the clustering and linked space within the app. A more experienced user may know exactly which space each variable is in before loading the app and so can pass them as additional parameters: \texttt{linked}, \texttt{flags} and \texttt{label}, keeping the clustering space in the \texttt{df} argument. An example would be using the included \texttt{Bikes} data:

\begin{verbatim}
pandemonium(df = Bikes$space1, linked = Bikes$space2, 
              getScore = outsideScore(Bikes$other$res, "Residual"))
\end{verbatim}

Here, the data includes original variables and latent variables from a single hidden-layer neural network model. In this case we use the model residuals as the scores; see the first case study for further information.

Additional arguments of the \texttt{pandemonium()} function can be used to provide user-defined computations for coordinates, score functions or dimension reduction methods, as explained in the section on modular inputs. The coordinate and score function can use the covariance matrix and a reference point in the clustering and linked space. These can be calculated within the app using \texttt{cov()} and \texttt{colMeans()} respectively, but alternative inputs can also be passed in as additional parameters when loading the app. If they are provided when loading the app, removing variables from the corresponding space will slice out that variable from the covariance matrix or reference point. Moving variables into a new space will result in \texttt{cov()} and \texttt{colMeans()} being used to recalculate these.

Once the app is loaded, the workflow can interactively be constructed by the user, depending on previous knowledge, research questions and intermediate results that are further explored. Example workflows are sketched in the case studies below.
To reproduce results outside the app, \texttt{makePlots()} is used to generate selected figures, and \texttt{writeResults()} allows for clustering results to be written to a CSV file. These can then be used for further analysis or as a saved result for comparison.

\section{Case studies}\label{case-studies}

\subsection{Exploring a Machine Learning Model}\label{exploring-a-machine-learning-model}

As a first application we explore the use of \CRANpkg{pandemonium} for understanding a machine learning model. In our example, we use a neural network to model the daily number of bikes rented in the bike sharing dataset from \citet{bikes}.

\begin{figure}

{\centering \includegraphics[width=0.7\linewidth]{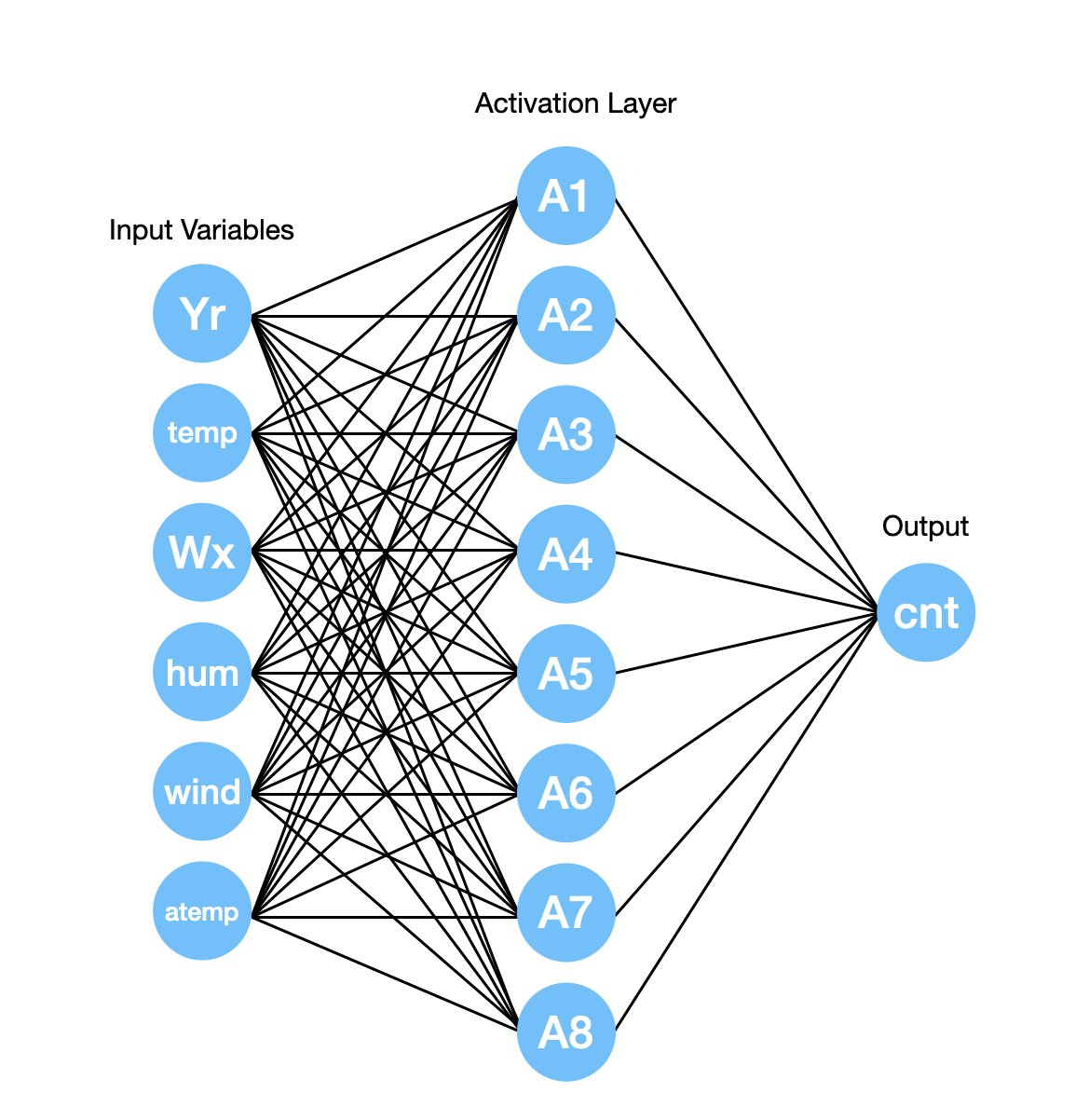} 

}

\caption{Diagram of the neural network used to model the daily number of bikes rented. The activations are used in the clustering space and the inputs in the linked space.}\label{fig:neural-network-im}
\end{figure}

We use six independent variables containing information about each day, including the weather conditions, wind and year. The response variable is the number of bike users recorded in the dataset. These six independent variables were used as predictors in a single hidden-layer neural network with eight hidden nodes and a ReLU activation function. The model was trained and used to predict bike rental counts with \CRANpkg{keras} \citep{keras}, as illustrated in Fig. \ref{fig:neural-network-im}. To reproduce the example, users can launch the app as shown in the usage section.

To explore the relation between predictors (the \(X_j\)s in the linked space), latent variables (the activations, which are the \(Y_i\)s in the clustering space) and model performance (for example the residuals for each observation, stored in the \(Z_k\)s), we structure all the information into a single dataset included in \CRANpkg{pandemonium}.
The model residuals in \(Z\) can be explored using the score function \texttt{outsideScore} in the pandemonium package with the \texttt{normCoords} coordinate function used for both \(X, Y\) spaces.

\begin{figure}
\includegraphics[width=0.5\linewidth]{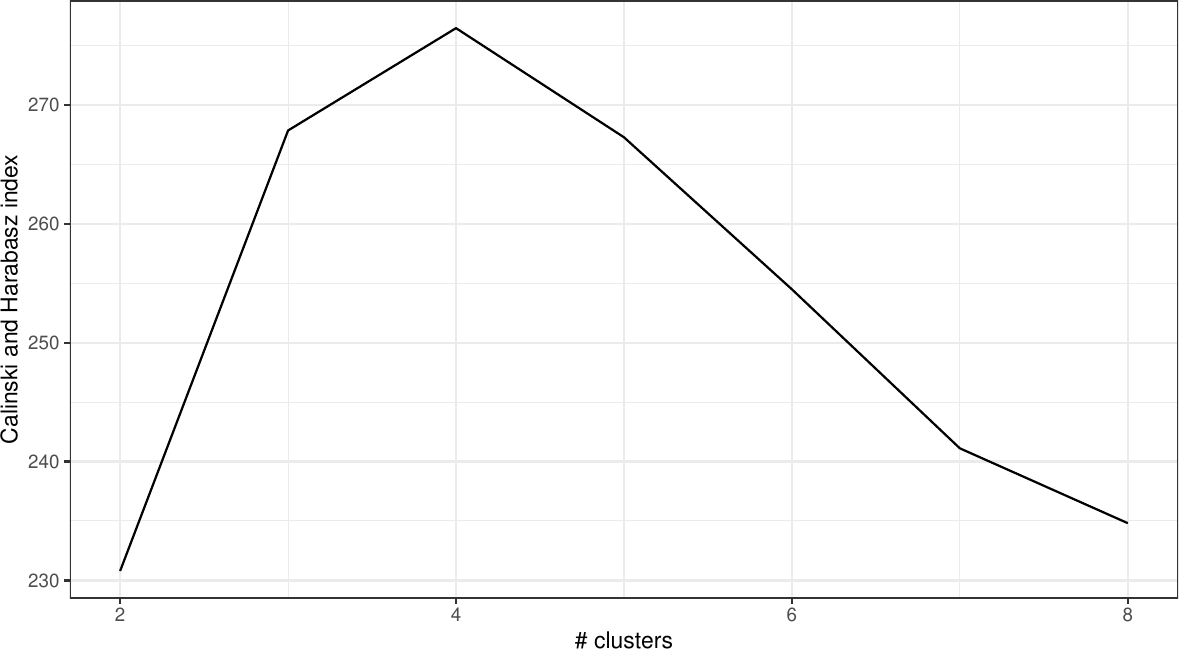} \includegraphics[width=0.5\linewidth]{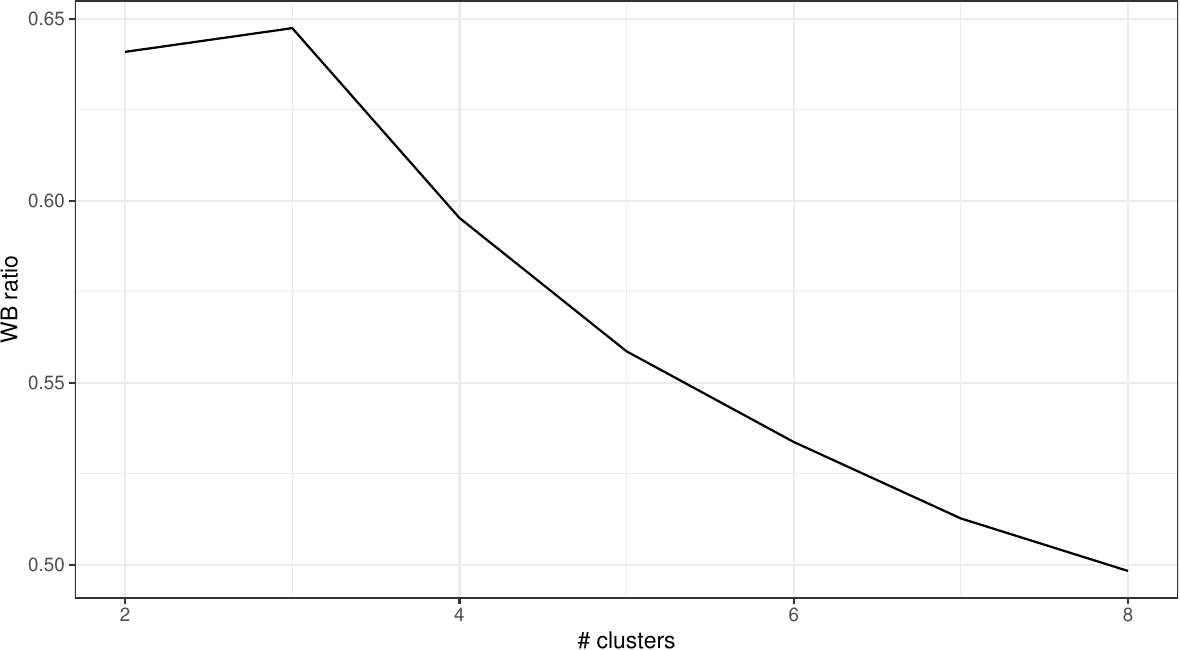} \caption{Selected panels from the statistics tab: The Calinski and Harabasz (CH) index (left), and the average within/between (WB) ratio (right).}\label{fig:bikes-stats}
\end{figure}

We begin by clustering the data with \texttt{ward.D2} linkage and Euclidean distances. The optimal number of clusters is selected by comparing the cluster statistics, see Fig. \ref{fig:bikes-stats}. The CH index indicates a four-cluster solution, while the WB ratio favours three clusters. In \CRANpkg{pandemonium}, the number of clusters can be adjusted interactively in the input tab, and compared across solutions in the comparison tab, enabling visual assessment to complement the statistical criteria when selecting the final number of clusters.

In this case, it is informative to link views obtained through non-linear dimension reduction using UMAP and linear projections visualised in a tour, as illustrated in Fig. \ref{fig:crosstalk-ex-pdf}. Since clusters are determined hierarchically, the four-cluster solution subdivides one of the clusters in the three-cluster solution. The UMAP display shows that the groupings identified through nonlinear dimension reduction differ from those suggested by the clustering, and this discrepancy can be further explored using the tour views. Examining the clustering space (the model activations) we clearly see the patterns introduced by the activation function, where points fall along lines or planes where some of the activations are close to zero. It is also clear that there are no well-separated clusters in this space. Interactive exploration through linked brushing shows that UMAP has grouped points that are more diffuse but with smooth transitions between our clusters, as well as points that fall along different linear patterns that are connected near the centre.

\begin{figure}
\includegraphics[width=0.5\linewidth]{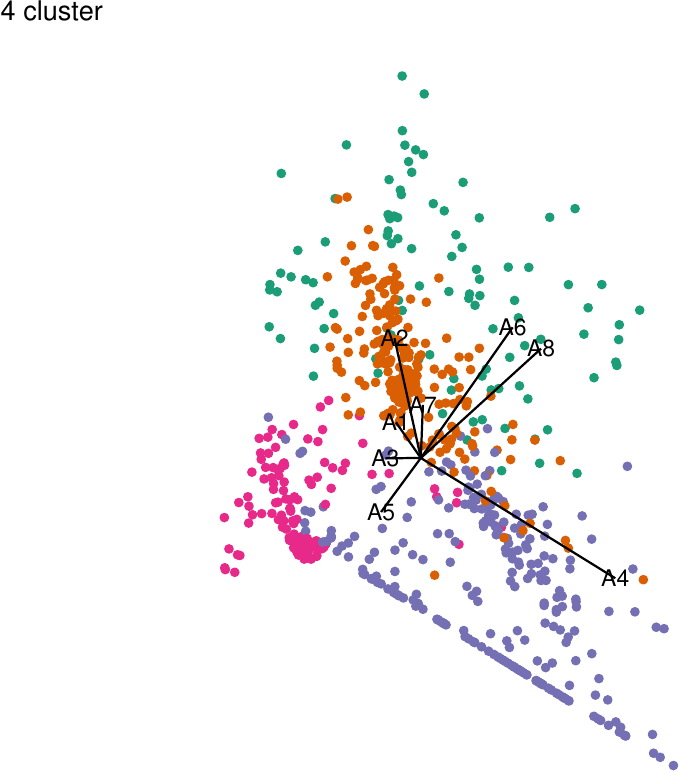} \includegraphics[width=0.5\linewidth]{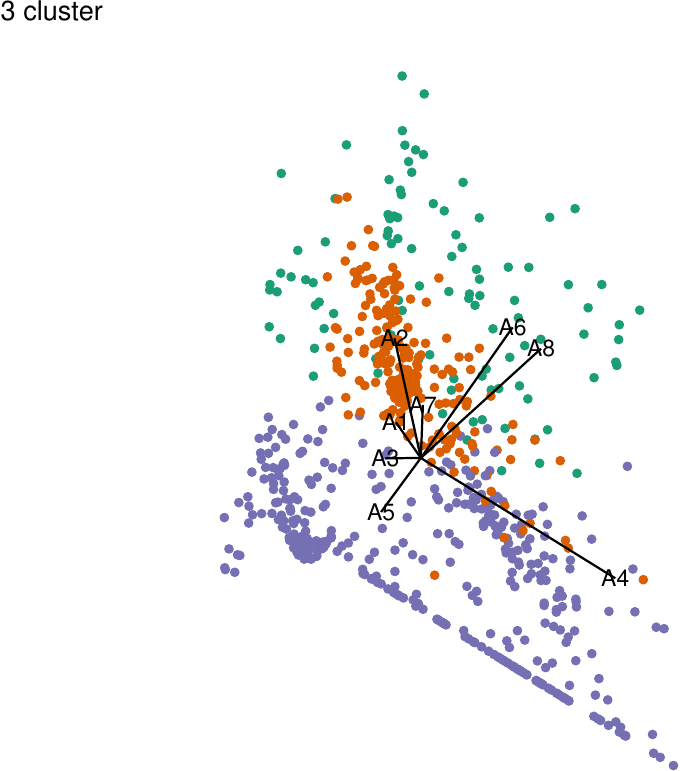} \caption{Final projection of LDA projection pursuit guided tour of bikes clustering space with 4 clusters, shown with colouring by 4 and 3 clusters. The purple cluster is extending along the A4 direction; the pink cluster appears to be sensitive to a combination of activations.}\label{fig:bikes-3vs4-tour}
\end{figure}

\begin{figure}

{\centering \includegraphics[width=0.9\linewidth]{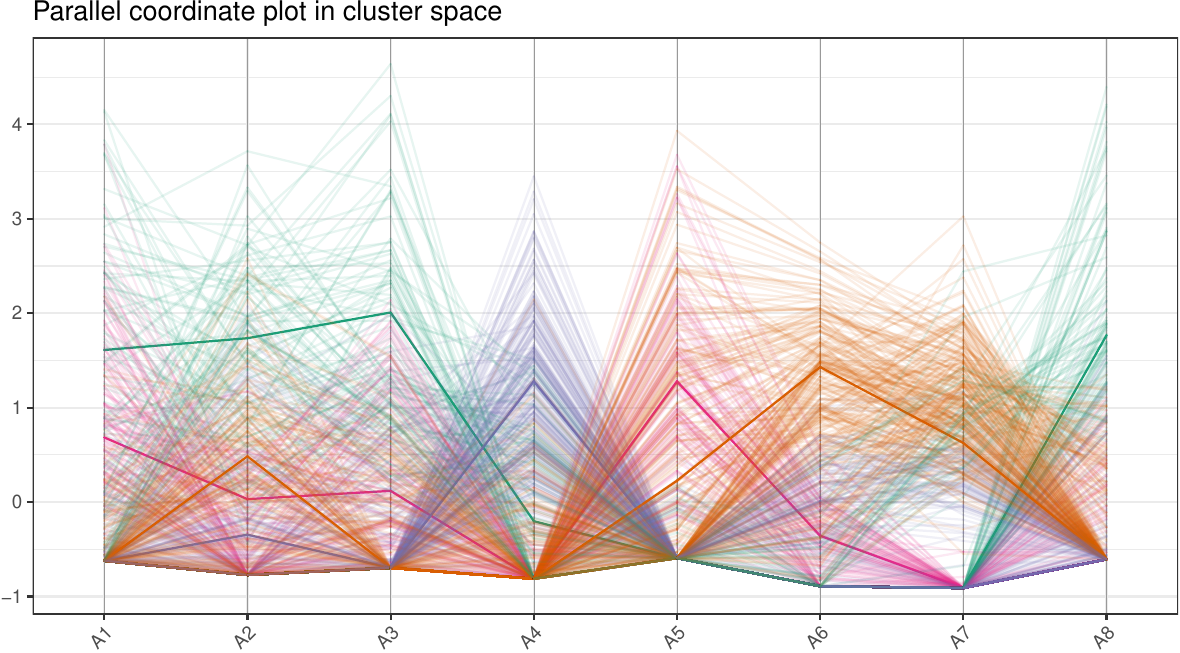} 

}

\caption{Parallel coordinate plot, where going from 3 to 4 clusters splits a large cluster into those shown in pink and purple, showing large differences in A4 and A5 direction. Highlighted lines correspond to cluster benchmarks.}\label{fig:bikes-pc}
\end{figure}

To better understand the differences between the three- and four-cluster solutions, we examine the final projection obtained from a guided tour using the LDA index to separate the four clusters, see Fig. \ref{fig:bikes-3vs4-tour}. By combining this information with what is seen in a parallel coordinate plot, we see that going from three to four clusters splits up components extending in the A4 vs A5 direction, see Fig. \ref{fig:bikes-pc}. Based on these observations, we proceed with the four-cluster solution for the subsequent analysis.

\begin{figure}

{\centering \includegraphics[width=0.6\linewidth]{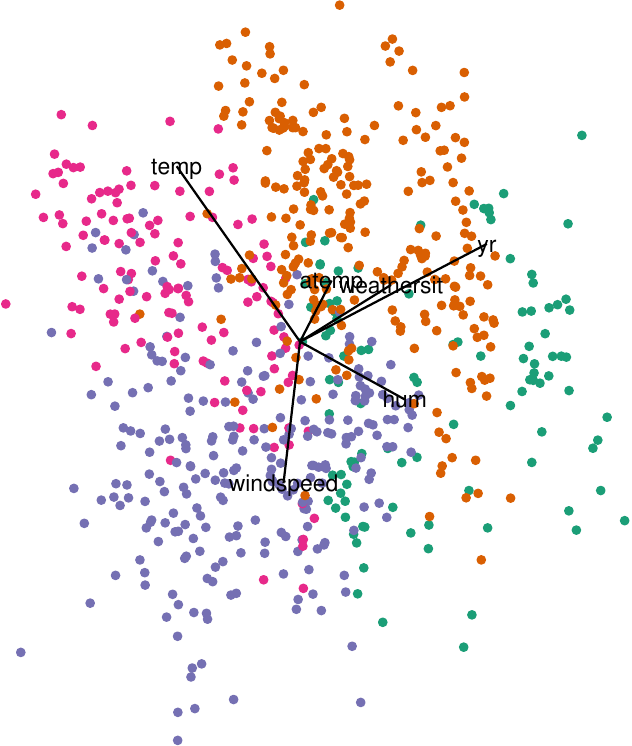} 

}

\caption{Final projection of LDA projection pursuit guided tour of bikes linked space with 4 clusters. The pink and purple clusters appear as separated along temperature and humidity inputs, and potentially also wind speed matters.}\label{fig:bikes-linked-tour}
\end{figure}

We next investigate what the cluster solution reveals about the connection between the clustering (activation) and linked (input) space. To begin, we again optimise the LDA index for cluster separation, but this time within the linked space, see Fig. \ref{fig:bikes-linked-tour}. From the final projection, we identify temperature and humidity as inputs of interest for separating the pink and purple clusters. The windspeed may also contribute to the cluster separation and could be further examined using a radial tour. Our next step is therefore to examine how the activation values of those identified in Fig. \ref{fig:bikes-3vs4-tour} vary across this two-dimensional input space.

\begin{figure}
\includegraphics[width=0.5\linewidth]{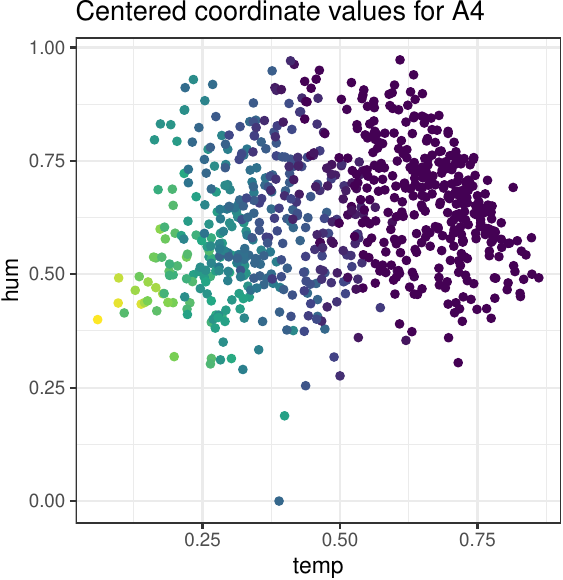} \includegraphics[width=0.5\linewidth]{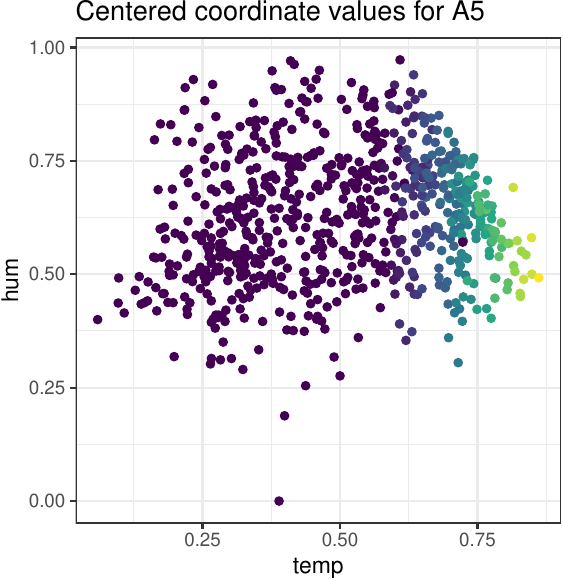} \includegraphics[width=0.5\linewidth]{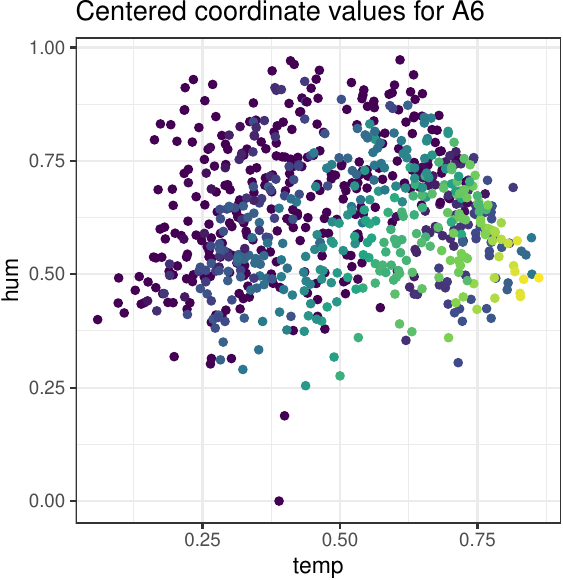} \includegraphics[width=0.5\linewidth]{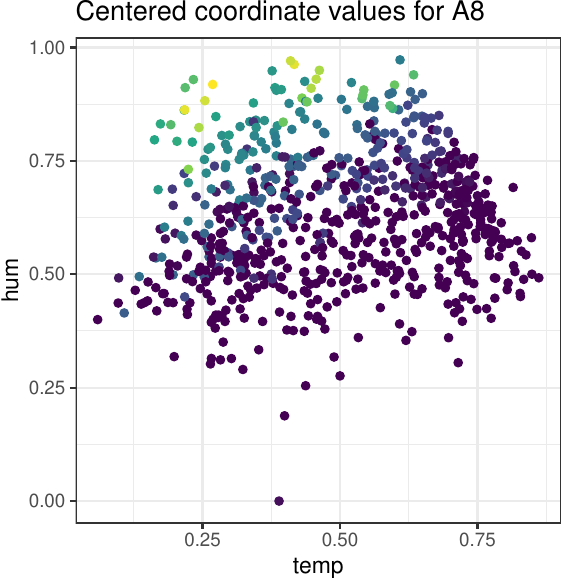} \caption{Centred coordinate plots for the four activations picked up by the guided tour, plotted on `temp` and `hum` variables.}\label{fig:bikes-obsplots}
\end{figure}

The relationships between each activation and the input parameters can be investigated through the centred coordinate plots. The coordinate values for each cluster variable are shown using a colour scale across a projection onto any two of the linked space (input parameters) variables. The plots for \texttt{A4}, \texttt{A5}, \texttt{A6} and \texttt{A8} are shown in Fig. \ref{fig:bikes-obsplots}. We can see that \texttt{A4} and \texttt{A5} have opposing patterns along the temperature direction, and that activations depend on humidity in relation to temperature values.

\begin{figure}

{\centering \includegraphics[width=1\linewidth]{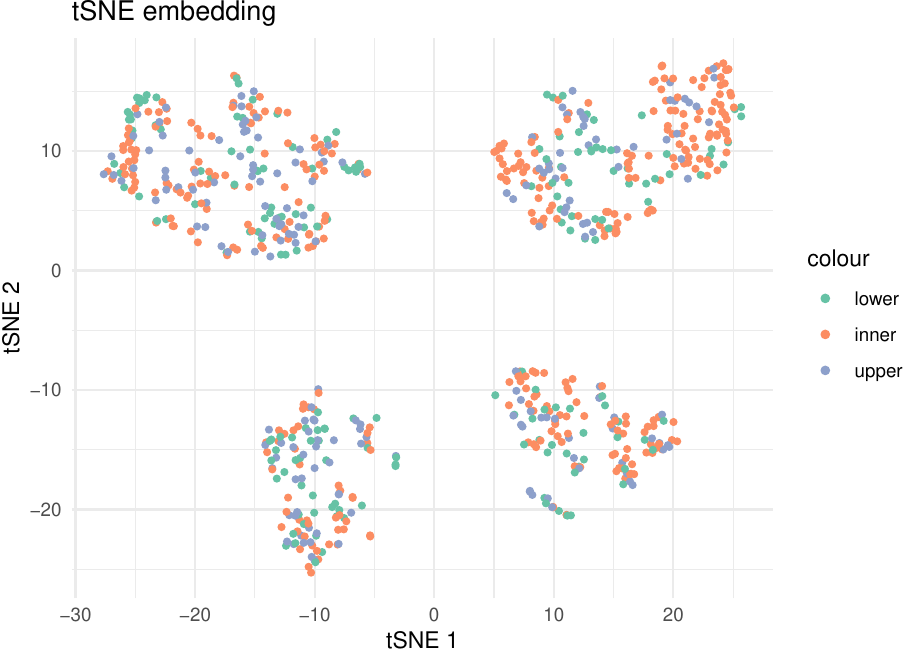} 

}

\caption{t-SNE reduction of linked space (input variables) with binned residual values for each point.}\label{fig:bikes-resbin-dimred}
\end{figure}

Finally, we examine the model residuals. These can be imported into \CRANpkg{pandemonium} using the \texttt{outsideScore} function. The scores can then be visualised in the dimension reduction and tour views either as a raw score gradient colouring or as binned score values. For illustration, the binned residuals are displayed in Fig. \ref{fig:bikes-resbin-dimred} on a t-SNE dimension reduction plot. No clear trends across the plot are apparent, suggesting that the model is not struggling in specific regions of the input space but fits well to the training data. Within \CRANpkg{pandemonium} we could further explore if there are patterns relating to clustering in either the activation or input spaces.

\subsection{Digging into a multi-parameter fit}\label{physics}

In this example, we illustrate the use of \CRANpkg{pandemonium} to study correlations and to identify important predictors in a high-dimensional model. The example is borrowed from particle physics, where a complex non-linear model with 150 parameters is used to fit certain experimental results and, in that way, extract information from the data \citep{LHCb:2024onj}. The dimensionality of a system with this many predictors and responses is too large to handle with \CRANpkg{pandemonium}, so the study is carried out in steps. In the first step, theoretical knowledge of the domain is used to infer smaller sets of potentially important predictors, which are then studied against a set of sixteen responses that are known to be important \citep{Capdevila:2018jhy} and have been previously measured by the same experimental collaboration \citep{LHCb:2020lmf}. This step is iterated, identifying all the important predictors which are then used to generate simplified models for further study \citep{McCoy2025}.

To illustrate the use of \CRANpkg{pandemonium} in this setting, we now illustrate its application to one of these small sets, consisting of six predictors (the \(X_j\)s in the linked space) and sixteen responses (the \(Y_i\)s in the clustering space). The focus here is on the exploration of the clustering result in the linked space, to understand which of the predictors have relevant effects on the responses, as well as correlations between the predictors.

The predictors chosen for this example include one that is known to be very important (\(X_1\)), one that is known to be much less important (\(X_2\)) and four whose importance we aim to infer (\(X_3\)-\(X_6\)). The data set is prepared by using the full model with 150 predictors to generate approximately 1000 sets of responses as follows. The space of six predictors is randomly sampled, using stratification to ensure the sample covers a range of five standard deviations around the best fit of \citet{LHCb:2024onj}. The sampling assumes that the six predictors follow a multivariate normal distribution centred at the best fit with a covariance matrix given by the corresponding six-dimensional submatrix obtained in the fit of \citet{LHCb:2024onj}. The remaining 144 predictors are fixed to their best-fit value.

We next use \CRANpkg{pandemonium} to cluster these models in response space and to explore the resulting structure. Following the discussion in \citet{Laa:2021dlg}, clustering for partitioning a continuous predictor space such as this one is based on the coordinate function
\begin{equation}
\widetilde{Y}_{kj} = \sum_{j'} \frac{1}{\sqrt{(\Sigma_Y^{-1})_{jj}}} (\Sigma_Y^{-1})_{jj'} (Y_{kj'} - Z_{j,\mathrm{exp}}),
\end{equation}
where the variance-covariance matrix \(\Sigma_Y\) in the clustering space encodes the correlated uncertainties arising from the experimental measurements. The reference point \(Z_{j,\mathrm{exp}}\) is taken to be the experimentally observed value (central value of the measurements in \citet{LHCb:2020lmf}), so that the theoretical predictions are compared directly to the measured data.
The computation of these coordinates is implemented in \CRANpkg{pandemonium}, in the function \texttt{pullCoords()}, which uses input provided in the \texttt{exp} argument of \texttt{pandemonium()} to set the reference point \(Z_{j,\mathrm{exp}}\). With these coordinates, using Euclidean distance mimics a \(\chi^2\) function \citep{Laa:2021dlg}\footnote{We also use Ward.D2 linkage for this example.}, leading to a statistical interpretation of the resulting clusters.

\begin{figure}
\includegraphics[width=0.5\linewidth]{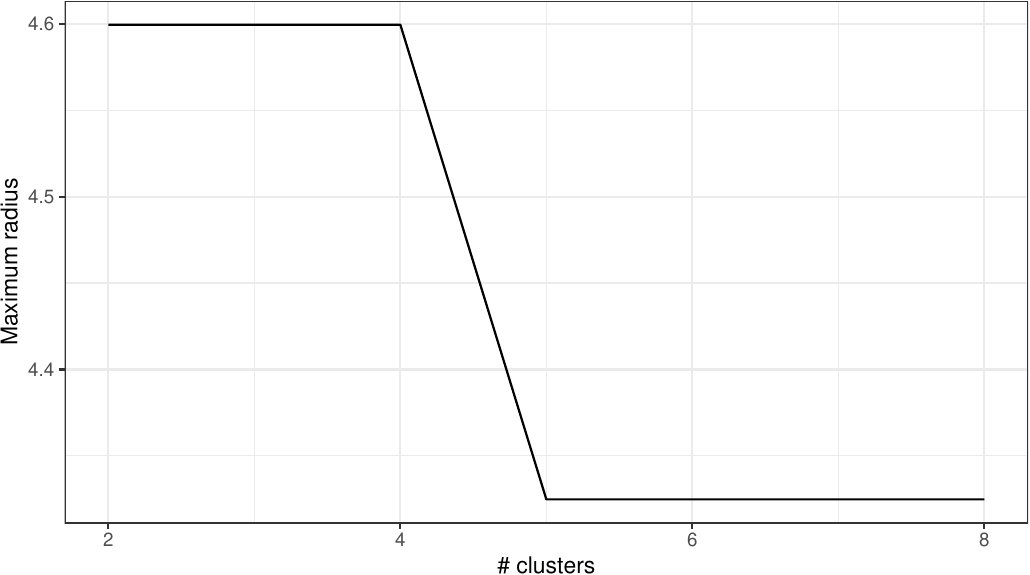} \includegraphics[width=0.5\linewidth]{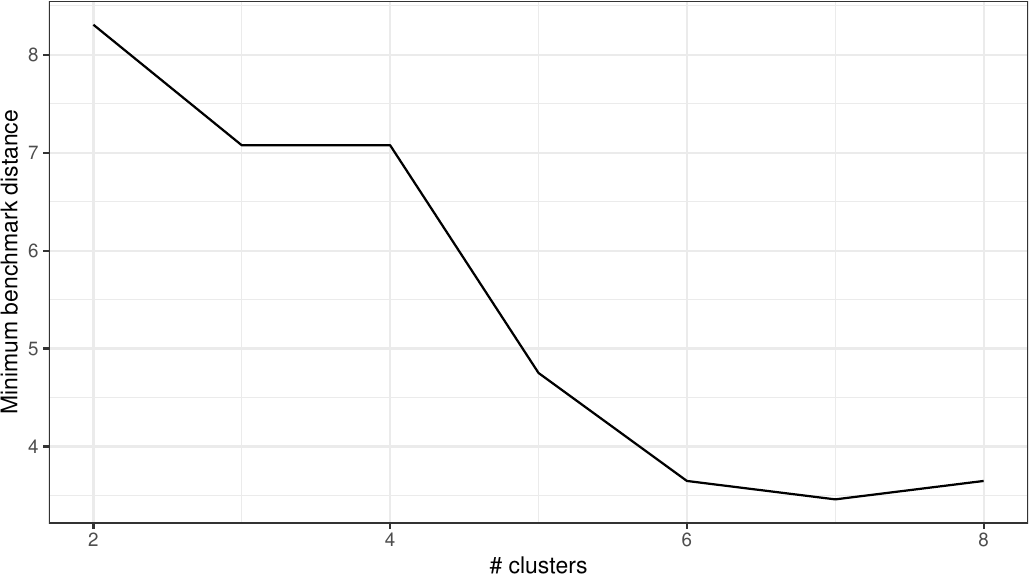} \caption{Selected panels from the statistics tab: the maximum cluster radius as a function of the number of clusters (left), and minimum distance between cluster benchmarks also as a function of number of clusters (right).}\label{fig:clusterstats}
\end{figure}

Given the statistical interpretation of the distance measure, we use the cluster radii and the minimum separation between cluster benchmarks to determine the appropriate number of clusters. At a chosen confidence level, all points within a cluster should be statistically indistinguishable, as quantified by the cluster radius, while different cluster benchmarks should remain sufficiently separated to be distinguishable from one another.
The results shown in Fig. \ref{fig:clusterstats} then suggest that we can use up to five clusters because this results in smaller cluster radii than minimum distance between cluster benchmarks. We choose four clusters for the remainder of this discussion.

\begin{figure}
\includegraphics[width=0.5\linewidth]{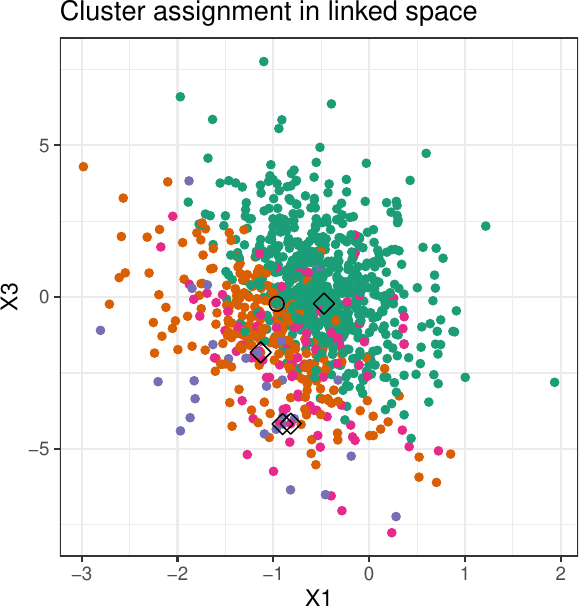} \includegraphics[width=0.5\linewidth]{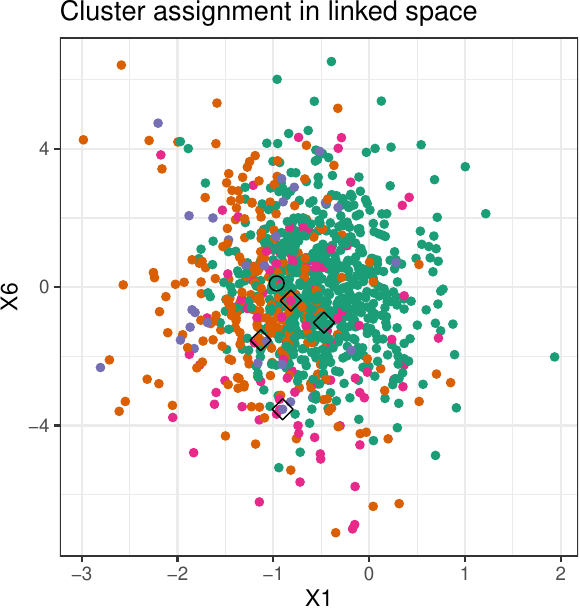} \caption{Four-six dimensional clusters in predictor space projected onto ($X_1$,$X_3$) (left panel) and onto ($X_1$,$X_6$) (right panel). We see a correlation pattern for cluster boundaries when $X_3$ is included, while there seems to be no dependence on $X_6$.}\label{fig:selected}
\end{figure}

A first step is to examine how the different predictors influence the clustering. Fig. \ref{fig:selected} shows projections of the clusters in predictor space onto the (\(X_1\)-\(X_3\)) and (\(X_1\)-\(X_6\)). These projections illustrate how \(X_3\) plays an important role in distinguishing the clusters, whereas \(X_6\) has little apparent impact (recall that \(X_1\) is known to be important from previous results).

\begin{figure}
\includegraphics[width=0.5\linewidth]{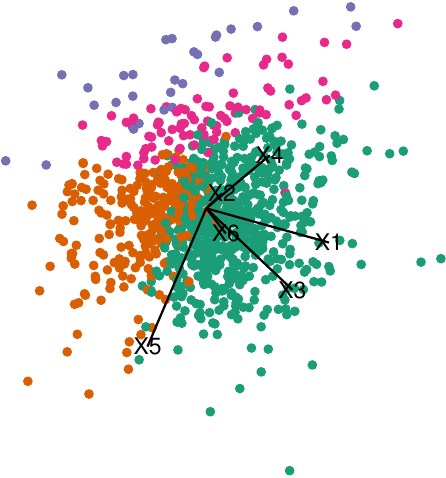} \includegraphics[width=0.5\linewidth]{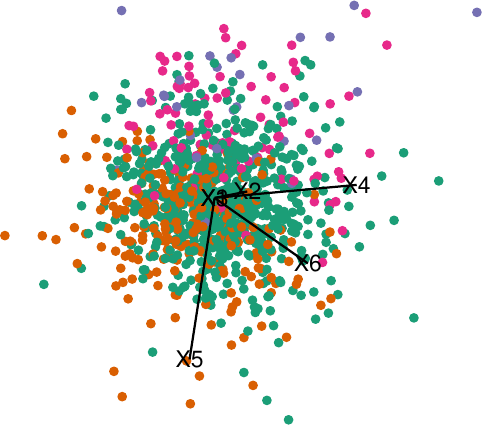} \caption{Results from a guided tour (left) and radial tour (right) exploring which predictors $X_i$ are relevant for the cluster separation. $X_1$ and $X_3$ appear as particularly important in the final view of the guided tour, and this is confirmed when removing them in a radial tour.}\label{fig:tour-finals}
\end{figure}

To confirm this conclusion, we inspect the result of running a guided tour with projection pursuit index PDA \citep{lee2010ppindex}. The result of this guided tour, shown in Fig. \ref{fig:tour-finals}, confirms that a combination of (\(X_1\), \(X_3\)) are mostly responsible for the observed clustering pattern. This conclusion is further supported by a radial tour \citep{2022arXiv221005228L} initialized from the PDA result and traversing the (\(X_1\), \(X_3\)) directions.

\section{Summary}\label{summary}

We have introduced the R package \CRANpkg{pandemonium}, which combines cluster analysis with linked visualisations to guide exploration in complementary spaces that carry different information about the same observations. The package clusters data in one variable space (typically the response space) to find groups with similar behaviour, then displays these clusters across both predictor and response spaces using non-linear dimension reduction and animated tours. Two examples demonstrate the package's usage, showing its flexibility across domains, from machine learning to physics, highlighting its value as an exploratory tool for understanding how predictor structures influence responses.

The package was designed to be accessible for novice R users who can perform most tasks in the graphical user interface. However, advanced R users will be able to take full advantage of the modularity implemented for various options in \CRANpkg{pandemonium}.
The implementation is allowing for modular inputs, such that the package can be applied to a range of different problems. We have demonstrated this for the physics example, and many additional applications can be considered. For example, we might have information about farmland both in terms of socio-economic factors gathered from farmers, as well as measurements of different soil properties, and want to explore relations between these, or compare information derived from satellite imagery with corresponding ground-based measurements.

\CRANpkg{pandemonium} is limited to mid-sized applications, both in terms of computing time and what can be usefully visualised. For large numbers of variables, we recommend variable selection or linear dimension reduction before interactive exploration. For large number of observations, options such as non-linear dimension reduction and the slice tour offer detailed visual exploration, but computing times can become an issue.
Another limitation arises from the trade-off between flexibility and simplicity. We have allowed modular inputs in places that were required for the applications we have encountered, but future applications might require an extension of the modularity. Similarly, in the visualisations, only some options can be changed interactively by the user.

A next step in further developing the implementation is to explore a broader range of applications in order to identify the limitations that should be addressed in future versions of the package.

\section{Acknowledgements}\label{acknowledgements}

The development of \CRANpkg{pandemonium} has been supported by Google Summer of Code 2025. We are grateful to Dianne Cook for helpful discussions.

\bibliography{RJreferences.bib}

\address{%
Gabriel McCoy\\
Monash University\\%
School of Physics and Astronomy\\ Melbourne, Australia\\
\textit{ORCiD: \href{https://orcid.org/0009-0008-3570-0361}{0009-0008-3570-0361}}\\%
\href{mailto:gabe.mccoy02@gmail.com}{\nolinkurl{gabe.mccoy02@gmail.com}}%
}

\address{%
German Valencia\\
Monash University\\%
School of Physics and Astronomy\\ Melbourne, Australia\\
\textit{ORCiD: \href{https://orcid.org/0000-0001-6600-1290}{0000-0001-6600-1290}}\\%
\href{mailto:german.valencia@monash.edu}{\nolinkurl{german.valencia@monash.edu}}%
}

\address{%
Ursula Laa\\
BOKU University\\%
Institute of Statistics\\ Vienna, Austria\\
\textit{ORCiD: \href{https://orcid.org/0000-0002-0249-6439}{0000-0002-0249-6439}}\\%
\href{mailto:ursula.laa@boku.ac.at}{\nolinkurl{ursula.laa@boku.ac.at}}%
}

%% file: RJreferences.bib
@article{McCoy2025,
   author =  {Gabriel McCoy and German Valencia},
   title = {},
   note = {Manuscript in preparation},
   year = {2025}
   }

@article{LHCb:2020lmf,
    author = "Aaij, Roel and others",
    collaboration = "LHCb",
    title = "{Measurement of $CP$-Averaged Observables in the $B^{0}\rightarrow K^{*0}\mu^{+}\mu^{-}$ Decay}",
    eprint = "2003.04831",
    archivePrefix = "arXiv",
    primaryClass = "hep-ex",
    reportNumber = "LHCb-PAPER-2020-002, CERN-EP-2020-027",
    doi = "10.1103/PhysRevLett.125.011802",
    journal = "Phys. Rev. Lett.",
    volume = "125",
    number = "1",
    pages = "011802",
    year = "2020"
}

@article{lee2010ppindex,
  title        = {A projection pursuit index for large p small n data},
  author       = {Lee, Eun Kyung and Cook, Dianne},
  journal      = {Statistics and Computing},
  volume       = {20},
  number       = {3},
  pages        = {381--392},
  year         = {2010},
  doi          = {10.1007/s11222-009-9131-1}
}

@Book{plotly,
    author = {Carson Sievert},
    title = {{Interactive Web-Based Data Visualizatio}n with R, plotly, and shiny},
    publisher = {Chapman and Hall/CRC},
    year = {2020},
    isbn = {9781138331457},
    url = {https://plotly-r.com},
  }

@Manual{crosstalk,
  title = {{crosstalk}: Inter-Widget Interactivity for HTML Widgets},
  author = {Joe Cheng and Carson Sievert},
  year = {2021},
  note = {R package version 1.1.1},
  url = {https://CRAN.R-project.org/package=crosstalk},
}

@Article{VIM,
  title = {Imputation with the {R} Package {VIM}},
  author = {Alexander Kowarik and Matthias Templ},
  journal = {Journal of Statistical Software},
  year = {2016},
  volume = {74},
  number = {7},
  pages = {1--16},
  doi = {10.18637/jss.v074.i07},
}

@Manual{alphahull,
  title = {alphahull: Generalization of the Convex Hull of a Sample of Points in the Plane},
  author = {Beatriz Pateiro-Lopez and Alberto Rodriguez-Casal and {.}},
  year = {2022},
  note = {R package version 2.5},
  url = {https://CRAN.R-project.org/package=alphahull},
  doi = {10.32614/CRAN.package.alphahull},
}

@Manual{detourr,
  title = {detourr: Portable and Performant Tour Animations},
  author = {Casper Hart and Earo Wang},
  year = {2025},
  note = {R package version 0.2.0},
  url = {https://CRAN.R-project.org/package=detourr},
  doi = {10.32614/CRAN.package.detourr},
}

@Manual{Rtsne1,
  title = {{Rtsne}: T-Distributed Stochastic Neighbor Embedding using Barnes-Hut Implementation},
  author = {Jesse H. Krijthe},
  year = {2015},
  note = {R package version 0.17},
  url = {https://github.com/jkrijthe/Rtsne},
}

@Article{Rtsne2,
  title = {Visualizing High-Dimensional Data Using t-SNE},
  volume = {9},
  pages = {2579-2605},
  year = {2008},
  author = {L.J.P. {van der Maaten} and G.E. Hinton},
  journal = {Journal of Machine Learning Research},
}

@Manual{uwot,
  title = {uwot: The Uniform Manifold Approximation and Projection (UMAP) Method for Dimensionality Reduction},
  author = {James Melville},
  year = {2025},
  note = {R package version 0.2.3},
  url = {https://CRAN.R-project.org/package=uwot},
  doi = {10.32614/CRAN.package.uwot},
}

@ARTICLE{2018arXiv180203426M,
       author = {{McInnes}, Leland and {Healy}, John and {Melville}, James},
        title = "{UMAP: Uniform Manifold Approximation and Projection for Dimension Reduction}",
      journal = {arXiv e-prints},
         year = 2018,
        month = feb,
          eid = {arXiv:1802.03426},
        pages = {arXiv:1802.03426},
          doi = {10.48550/arXiv.1802.03426},
archivePrefix = {arXiv},
       eprint = {1802.03426},
 primaryClass = {stat.ML},
       adsurl = {https://ui.adsabs.harvard.edu/abs/2018arXiv180203426M}
}

@Article{tourr,
  title = {{tourr}: An {R} Package for Exploring Multivariate Data with Projections},
  author = {Hadley Wickham and Dianne Cook and Heike Hofmann and Andreas Buja},
  journal = {Journal of Statistical Software},
  year = {2011},
  volume = {40},
  number = {2},
  pages = {1--18},
  url = {https://doi.org/10.18637/jss.v040.i02},
}

@Article{tourr-slice,
  title = {A slice tour for finding hollowness in high-dimensional data},
  author = {Ursula Laa and Dianne Cook and German Valencia},
  journal = {Journal of Computational and Graphical Statistics},
  year = {2020},
  volume = {29},
  number = {3},
  pages = {681--687},
  url = {https://doi.org/10.1080/10618600.2020.1777140},
}

@article{Laa:2021dlg,
    author = "Laa, Ursula and Valencia, German",
    title = "{Pandemonium: a clustering tool to partition parameter space{\textemdash}application to the B anomalies}",
    eprint = "2103.07937",
    archivePrefix = "arXiv",
    primaryClass = "physics.data-an",
    doi = "10.1140/epjp/s13360-021-02310-1",
    journal = "Eur. Phys. J. Plus",
    volume = "137",
    number = "1",
    pages = "145",
    year = "2022"
}

@article{Capdevila:2018jhy,
    author = "Capdevila, Bernat and Laa, Ursula and Valencia, German",
    title = "{Anatomy of a six-parameter fit to the $b\to s \ell^+\ell^-$ anomalies}",
    eprint = "1811.10793",
    archivePrefix = "arXiv",
    primaryClass = "hep-ph",
    doi = "10.1140/epjc/s10052-019-6944-8",
    journal = "Eur. Phys. J. C",
    volume = "79",
    number = "6",
    pages = "462",
    year = "2019"
}

@article{LHCb:2024onj,
    author = "Aaij, Roel and others",
    collaboration = "LHCb",
    title = "{Comprehensive analysis of local and nonlocal amplitudes in the B$^{0}${\textrightarrow} K$^{*0}${\ensuremath{\mu}}$^{+}${\ensuremath{\mu}}$^{-}$ decay}",
    eprint = "2405.17347",
    archivePrefix = "arXiv",
    primaryClass = "hep-ex",
    reportNumber = "LHCb-PAPER-2024-011, CERN-EP-2024-122",
    doi = "10.1007/JHEP09(2024)026",
    journal = "JHEP",
    volume = "09",
    pages = "026",
    year = "2024"
}

@article{vpse,
	author = {Piccolotto, Nikolaus and B{\"o}gl, Markus and Miksch, Silvia},
	journal = {Computer Graphics Forum},
	number = {6},
	pages = {e14785},
	title = {Visual Parameter Space Exploration in Time and Space},
	volume = {42},
	year = {2023}}

@Manual{shiny,
    title = {shiny: Web Application Framework for R},
    author = {Winston Chang and Joe Cheng and JJ Allaire and Carson Sievert and Barret Schloerke and Yihui Xie and Jeff Allen and Jonathan McPherson and Alan Dipert and Barbara Borges},
    year = {2025},
    note = {R package version 1.11.1},
    url = {https://CRAN.R-project.org/package=shiny},
    doi = {10.32614/CRAN.package.shiny},
  }

@article{JSSv076i10,
	author = {Sieger, Tom{\'a}{\v s} and Hurley, Catherine B. and Fi{\v s}er, Karel and Beleites, Claudia},
	journal = {Journal of Statistical Software},
	number = {10},
	pages = {1--22},
	title = {Interactive Dendrograms: The R Packages idendro and idendr0},
	volume = {76},
	year = {2017}}

@ARTICLE{drtool,
       author = {{Lin}, Justin and {Fukuyama}, Julia},
        title = "{DRtool: An Interactive Tool for Analyzing High-Dimensional Clusterings}",
      journal = {arXiv e-prints},
         year = 2025,
        month = sep,
          eid = {arXiv:2509.04603},
        pages = {arXiv:2509.04603},
          doi = {10.48550/arXiv.2509.04603},
archivePrefix = {arXiv},
       eprint = {2509.04603},
 primaryClass = {stat.AP},
       adsurl = {https://ui.adsabs.harvard.edu/abs/2025arXiv250904603L}
}

@article{lionfish,
title={Demonstrating the Capabilities of the lionfish Software for Interactive Visualization of Market Segmentation Partitions}, volume={54}, url={https://ajs.or.at/index.php/ajs/article/view/2058}, DOI={10.17713/ajs.v54i3.2058},
number={3}, journal={Austrian Journal of Statistics}, author={Medl, Matthias and Cook, Dianne and Laa, Ursula}, year={2025}, month={Apr.}, pages={71–99} }

@Article{dendextend,
    author = {Tal Galili},
    title = {dendextend: an R package for visualizing, adjusting, and comparing trees of hierarchical clustering},
    journal = {Bioinformatics},
    year = {2015},
    doi = {10.1093/bioinformatics/btv428},
    url = {https://doi.org/10.1093/bioinformatics/btv428},
    eprint = {https://academic.oup.com/bioinformatics/article-pdf/31/22/3718/17122682/btv428.pdf},
  }

@Manual{fpc,
    title = {fpc: Flexible Procedures for Clustering},
    author = {Christian Hennig},
    year = {2024},
    note = {R package version 2.2-13},
    url = {https://CRAN.R-project.org/package=fpc},
    doi = {10.32614/CRAN.package.fpc},
  }

@article{bikes,
year={2013}, 
issn={2192-6352}, 
journal={Progress in Artificial Intelligence},
doi={10.1007/s13748-013-0040-3},
url = {https://doi.org/10.1007/s13748-013-0040-3},
title={Event labeling combining ensemble detectors and background knowledge},
publisher={Springer Berlin Heidelberg}, 
author={Fanaee-T, Hadi and Gama, Joao}, 
pages={1-15}  
}

@Manual{keras,
    title = {keras: R Interface to 'Keras'},
    author = {JJ Allaire and François Chollet},
    year = {2024},
    note = {R package version 2.15.0},
    url = {https://CRAN.R-project.org/package=keras},
    doi = {10.32614/CRAN.package.keras},
  }

@Manual{htmlwidgets,
    title = {htmlwidgets: HTML Widgets for R},
    author = {Ramnath Vaidyanathan and Yihui Xie and JJ Allaire and Joe Cheng and Carson Sievert and Kenton Russell},
    year = {2023},
    note = {R package version 1.6.4},
    url = {https://CRAN.R-project.org/package=htmlwidgets},
    doi = {10.32614/CRAN.package.htmlwidgets},
  }

@article{ROUSSEEUW198753,
	author = {Peter J. Rousseeuw},
	journal = {Journal of Computational and Applied Mathematics},
	pages = {53-65},
	title = {Silhouettes: A graphical aid to the interpretation and validation of cluster analysis},
	volume = {20},
	year = {1987}}

@article{Lee_Laa_Cook_2022,
place={The Netherlands},
title={Casting multiple shadows: interactive data visualisation with tours and embeddings},
volume={2},
url={https://jdssv.org/index.php/jdssv/article/view/21},
DOI={10.52933/jdssv.v2i3.21},
author={Lee, Stuart and Laa, Ursula and Cook, Dianne},
year={2022}, month={May} }

@article{RJ-2023-052,
  author = {Hart, Casper and Wang, Earo},
  title = {Taking the Scenic Route: Interactive and Performant Tour Animations},
  journal = {The R Journal},
  year = {2023},
  note = {https://doi.org/10.32614/RJ-2023-052},
  doi = {10.32614/RJ-2023-052},
  volume = {15},
  issue = {2},
  issn = {2073-4859},
  pages = {307-329}
}

@article{CBCH94, 
        Author = {Cook, D. and Buja, A. and Cabrera, J. and Hurley, C.},
        Journal = {Journal of Computational and Graphical Statistics},
        Number = 3,
        Pages = {155--172},
        Title = {Grand Tour and Projection Pursuit},
        Volume = 4,
        Year = 1995}

@article{2022arXiv221005228L,
        author = {Ursula Laa and Alex Aumann and Dianne Cook and German Valencia},
        journal = {Journal of Computational and Graphical Statistics},
        number = {3},
        pages = {1229-1236},
        title = {New and Simplified Manual Controls for Projection and Slice Tours, With Application to Exploring Classification Boundaries in High Dimensions},
        volume = {32},
        url = {https://doi.org/10.1080/10618600.2023.2206459},
        year = {2023}}

@article{tourreview,
author = {Lee, Stuart and Cook, Dianne and da Silva, Natalia and Laa, Ursula and Spyrison, Nicholas and Wang, Earo and Zhang, H. Sherry},
title = {The state-of-the-art on tours for dynamic visualization of high-dimensional data},
journal = {WIREs Computational Statistics},
volume = {14},
number = {4},
pages = {e1573},
doi = {https://doi.org/10.1002/wics.1573},
url = {https://wires.onlinelibrary.wiley.com/doi/abs/10.1002/wics.1573},
eprint = {https://wires.onlinelibrary.wiley.com/doi/pdf/10.1002/wics.1573},
year = {2022}
}
